\documentclass[%
  reprint,
  showpacs,preprintnumbers,
  amsmath,amssymb,
  aps,
prl,
]{revtex4-1}

\usepackage{graphicx}% Include figure files
\usepackage{dcolumn}% Align table columns on decimal point
\usepackage{bm}% bold math
\usepackage{hyperref}% add hypertext capabilities
\usepackage{color}
\usepackage{ulem}

\usepackage{times}
\usepackage{booktabs}

\begin{document}

\preprint{APS/123-QED}

\title{
N\'eel- and Bloch-Type Magnetic Vortices in Rashba Metals
}% Force line breaks with \\

\author{Satoru Hayami$^1$ and Yukitoshi Motome$^2$}
\affiliation{%
 $^1$Department of Physics, Hokkaido University, Sapporo 060-0810, Japan \\
 $^2$Department of Applied Physics, University of Tokyo, Tokyo 113-8656, Japan
}%
 
\begin{abstract}
We theoretically study noncoplanar spin textures in polar magnetic conductors. 
Starting from the Kondo lattice model with the Rashba spin-orbit coupling, we derive an effective spin model with generalized Ruderman-Kittel-Kasuya-Yosida interactions including the anisotropic and antisymmetric exchange interactions. 
By performing simulated annealing for the effective model, we find that a vortex crystal of N\'eel type is stabilized even in the absence of a magnetic field. 
Moreover, we demonstrate that a Bloch-type vortex crystal, which is usually associated with the Dresselhaus spin-orbit coupling, can also be realized in our Rashba-based model. 
A magnetic field turns the vortex crystals into N\'eel- and Bloch-type skyrmion-like crystals. 
Our results underscore that the interplay between the spin-orbit coupling and itinerant magnetism brings fertile possibilities of noncoplanar magnetic orderings. 
\end{abstract}
\maketitle

Noncoplanar magnetic textures have attracted great interest in condensed matter physics, as they often give rise to topologically nontrivial quantum states and related fascinating phenomena. 
The noncoplanar orders can simultaneously activate the scalar chirality, which is denoted as a scalar triple product of spins $\mathbf{S}_i \cdot (\mathbf{S}_j \times \mathbf{S}_k)$, in addition to the primary magnetic order parameter. 
The scalar chirality generates an emergent magnetic field for itinerant electrons through the spin Berry phase mechanism~\cite{berry1984quantal}, and hence, has great potential to induce and control unconventional quantum phenomena, such as the anomalous Hall effect~\cite{Loss_PhysRevB.45.13544,Ye_PhysRevLett.83.3737,Ohgushi_PhysRevB.62.R6065,tatara2002chirality}. 

\begin{figure}[t]
\begin{center}
\includegraphics[width=1.0 \hsize]{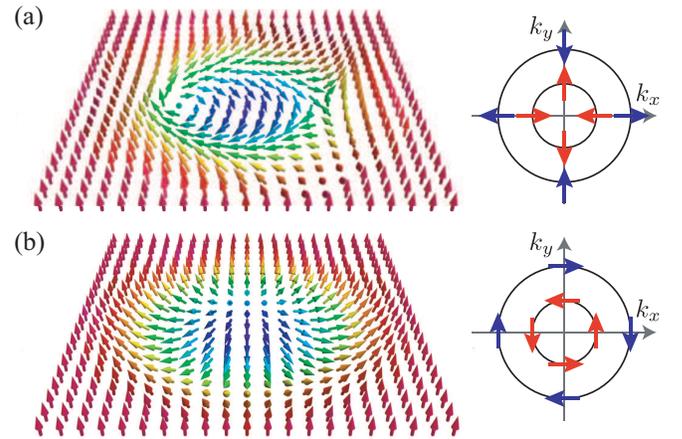} 
\caption{
\label{Fig:ponti}
Schematic pictures of (a) Bloch-type and (b) N\'eel-type spin textures. 
The right panels show the schematic Fermi surfaces split by the SOC favoring each skrymion texture: (a) Dresselhaus type and (b) Rashba type. 
The arrows represent the spin polarization on the Fermi surfaces. 
}
\end{center}
\end{figure}

Such noncoplanar orders have been extensively studied in noncentrosymmetric systems where the spin-orbit coupling (SOC) plays an important role. 
For instance, hexagonal skyrmion crystals with noncoplanar spin textures emerge in noncentrosymmetric systems in an applied magnetic field~\cite{Bogdanov89,rossler2006spontaneous,nagaosa2013topological}. 
The skyrmion crystals are classified according to their vorticity and helicity. 
For instance, a Bloch-type skyrmion with the helicity $\pm\pi/2$ can be stabilized in chiral magnets with the Dresselhaus SOC~\cite{Muhlbauer_2009skyrmion,yu2010real,seki2012observation} [Fig.~\ref{Fig:ponti}(a)], while a N\'eel-type one with the helicity $0$ or $\pi$ may appear in polar magnets with the Rashba SOC~\cite{kezsmarki_neel-type_2015,Kurumaji_PhysRevLett.119.237201} [Fig.~\ref{Fig:ponti}(b)]. 

Besides the SOC, recent theoretical studies unveiled alternative origins of similar noncoplanar spin textures: e.g., frustrated exchange interactions and dipolar interactions in localized spin systems~\cite{yu2012magnetic,Okubo_PhysRevLett.108.017206,Kamiya_PhysRevX.4.011023,Marmorini2014,leonov2015multiply,Lin_PhysRevB.93.064430,Hayami_PhysRevB.93.184413,Batista2016}, and the spin-charge coupling between itinerant electron spins and localized spins~\cite{Martin_PhysRevLett.101.156402,Akagi_JPSJ.79.083711,Kumar_PhysRevLett.105.216405,Chern_PhysRevLett.105.226403,Kato_PhysRevLett.105.266405,Hayami_PhysRevB.89.085124,Hayami_PhysRevB.90.060402,Venderbos_PhysRevB.93.115108,Hayami_PhysRevB.94.024424,Ozawa_doi:10.7566/JPSJ.85.103703,Ozawa_PhysRevLett.118.147205,Hayami_PhysRevB.95.224424}. 
These mechanisms would be useful not only to clarify an unknown origin of noncoplanar orderings, e.g., in SrFeO$_3$~\cite{Ishiwata_PhysRevB.84.054427}, MnSc$_2$S$_4$~\cite{Gao2016Spiral}, and CeAuSb$_2$~\cite{marcus2017multi}, but also to provide a reference for exploring further exotic spin textures. 

In the present study, we push forward these theoretical studies to a more realistic situation, by taking into account both the SOC and the spin-charge coupling in magnetic conductors. 
Our aim is to illuminate noncoplanar spin textures induced by the interplay between the two couplings. 
To this end, starting from the Kondo lattice model with the SOC, we derive an effective spin model with anisotropic and antisymmetric exchange interactions in momentum space. 
We find that the effective model exhibits a stable N\'eel-type vortex crystal (VC) even in the absence of a magnetic field. 
Moreover, we show that a Bloch-type VC, which is usually induced by the Dresselhaus SOC, can also be realized within the same model.  
We also discuss how these VCs evolve to skyrmion-like crystals and exhibit phase transitions to other spin states in an applied magnetic field. 

Let us consider the Kondo lattice model with the SOC on a square lattice with the point group $C_{4v}$. 
The following analysis can be extended to any polar and chiral systems. 
The Hamiltonian is given by 
\begin{eqnarray}
\label{eq:Ham_krep}
\mathcal{H}=& &\sum_{\mathbf{k} \sigma} (\varepsilon_\mathbf{k}-\mu) c^{\dagger}_{\mathbf{k}\sigma}c_{\mathbf{k}\sigma} 
+ J_{\rm K} \sum_{\mathbf{k}\mathbf{q}\sigma\sigma'}
c^{\dagger}_{\mathbf{k}\sigma}\bm{\sigma}_{\sigma \sigma'}c_{\mathbf{k}+\mathbf{q}\sigma'} \cdot \mathbf{S}_{\mathbf{q}} \nonumber \\
&+&  \sum_{\mathbf{k}} \mathbf{g}_{\mathbf{k}} \cdot c^{\dagger}_{\mathbf{k}\sigma}\bm{\sigma}_{\sigma \sigma'}c_{\mathbf{k}\sigma'},  
\end{eqnarray}
where $c^{\dagger}_{\mathbf{k}\sigma}$ ($c_{\mathbf{k}\sigma}$) is a creation (annihilation) operator of an itinerant electron at wave number $\mathbf{k}$ and spin $\sigma$. 
The first term in Eq.~(\ref{eq:Ham_krep}) represents the kinetic motion of itinerant electrons; $\varepsilon_\mathbf{k}$ is the energy dispersion of free electrons and $\mu$ is the chemical potential. 
The second term represents the Kondo coupling between itinerant electron spins and localized spins; $\bm{\sigma}=(\sigma^x,\sigma^y,\sigma^z)$ is the vector of Pauli matrices, $\mathbf{S}_{\mathbf{q}}$ is the Fourier transform of a localized spin $\mathbf{S}_i$ defined at site $i$ with the fixed length $|\mathbf{S}_i|=1$, and $J_{\rm K}$ is the exchange coupling constant. 
The sign of $J_{\rm K}$ is irrelevant, as we regard the localized spins as classical ones for simplicity. 
The third term in Eq.~(\ref{eq:Ham_krep}) represents the SOC. 
We here consider the Rashba-type SOC with polar systems in mind: $\mathbf{g}_{\mathbf{k}} =(g_{\mathbf{k}}^x, g_{\mathbf{k}}^y) \propto (\sin k_y,-\sin k_x)$, which is induced by the mirror symmetry breaking with respect to the square plane. 
Related models were studied in the context of an electron gas coupled with magnetic impurities~\cite{Imamura_PhysRevB.69.121303,malecki2007two,Schulz_PhysRevB.79.205432,Chesi_PhysRevB.82.165303,shibuya2016magnetic,bouaziz2017chiral}. 

Instead of directly studying the ground state of the model in Eq.~(\ref{eq:Ham_krep}), which requires laborious computational calculations, we here extract important magnetic contributions by deriving an effective spin model as follows. 
First, we assume that $J_{\rm K}$ is small enough compared to the bandwidth of itinerant electrons.  
In the weak-coupling regime with $\mathbf{g}_{\mathbf{k}}=\mathbf{0}$, the effective magnetic interaction between localized spins mediated by itinerant electrons is described by  the Ruderman-Kittel-Kasuya-Yosida (RKKY) interaction, which is obtained by the second-order perturbation with respect to $J_{\rm K}$~\cite{Ruderman,Kasuya,Yosida1957}. 
Extending the analysis to the model in Eq.~(\ref{eq:Ham_krep}) by including the SOC term in the unperturbed Hamiltonian, we obtain 
\begin{eqnarray}
\label{eq:RKKYHam}
\mathcal{H} =-\frac{J_{\rm K}^2}{4N}  \sum_{\mathbf{k} \mathbf{q} }  \sum_{\sigma \sigma'}
\chi^0_{\mathbf{k}\sigma\sigma'}
\left(\mathcal{H}^{\rm iso}+\mathcal{H}^{\rm ani}+\mathcal{H}^{\rm DM}
\right),  
\end{eqnarray}
where 
\begin{eqnarray}
  \label{eq:RKKYHam_iso}
\mathcal{H}^{\rm iso} &=& (I_{\sigma \sigma'}+\tau^x_{\sigma \sigma'}) \, 
\mathbf{S}_{\mathbf{q}}\cdot \mathbf{S}_{-\mathbf{q}}, \\
\label{eq:RKKYHam_ani}
\mathcal{H}^{\rm ani}&=&  (I_{\sigma \sigma'}-\tau^x_{\sigma \sigma'}) 
\sum_{\nu \neq \nu' } 
\Big[ \left(\tilde{g}^{\nu}_{\mathbf{k}}\tilde{g}^{\nu'}_{\mathbf{k}+\mathbf{q}}
S^{\nu}_{\mathbf{q}} S^{\nu'}_{-\mathbf{q}} 
+{\rm H.c.}\right) \nonumber  \\
& 
\qquad &+\tilde{g}^\nu_{\mathbf{k}} \tilde{g}^\nu_{\mathbf{k}+\mathbf{q}} 
 (S^\nu_{\mathbf{q}} S^\nu_{-\mathbf{q}}-S^{\nu'}_{\mathbf{q}} S^{\nu'}_{-\mathbf{q}}-S^z_{\mathbf{q}} S^z_{-\mathbf{q}}) \Big],\\
\label{eq:RKKYHam_DM}
\mathcal{H}^{{\rm DM}}&=&   \sum_{\nu}
(i\tau^z_{\sigma \sigma'}G^\nu_{\mathbf{k}\mathbf{q}-}
    - \tau^y_{\sigma \sigma'}G^\nu_{\mathbf{k}\mathbf{q}+}) (\mathbf{S}_{\mathbf{q}}\times \mathbf{S}_{-\mathbf{q}})^\nu. 
\end{eqnarray}
Here, $\mathcal{H}^{{\rm iso}}$ desrcibes the conventional RKKY interaction, which is isotropic in spin space, while 
$\mathcal{H}^{{\rm ani}}$ and $\mathcal{H}^{{\rm DM}}$ are additional anisotropic contributions arising from the SOC. 
$\mathcal{H}^{{\rm ani}}$ is the anisotropic exchange interaction, and $\mathcal{H}^{{\rm DM}}$ is the antisymmetric exchange interaction, corresponding to the Dzyaloshinskii-Moriya (DM) interaction~\cite{dzyaloshinsky1958thermodynamic,moriya1960anisotropic}. 
In Eq.~(\ref{eq:RKKYHam}), $\chi^0_{\mathbf{k}\sigma\sigma'}=[f(E_{\mathbf{k}\sigma})-f(E_{\mathbf{k}+\mathbf{q}\sigma'})]/(E_{\mathbf{k}+\mathbf{q}\sigma'}-E_{\mathbf{k}\sigma})$ represents the spin-dependent bare susceptibility of itinerant electrons at the wave number $\mathbf{k}$, where $E_{\mathbf{k}\sigma}=\varepsilon_{\mathbf{k}}-\mu+ \sigma\sqrt{(g^x_{\mathbf{k}})^2+(g^y_{\mathbf{k}})^2}$ and $f(E_{\mathbf{k}\sigma})$ is the Fermi distribution function. 
In Eqs.~(\ref{eq:RKKYHam_iso})-(\ref{eq:RKKYHam_DM}), $I_{\sigma \sigma'}$ and $\tau^{\mu}_{\sigma \sigma'}$ ($\mu=x,y,z$) are the identity and Pauli matrices in pseudospin space, respectively, $\tilde{g}^\nu_{\mathbf{k}}=g^\nu_{\mathbf{k}}/|\mathbf{g}_{\mathbf{k}}|$, $G^\nu_{\mathbf{k}\mathbf{q}\pm}=\tilde{g}^\nu_{\mathbf{k}+\mathbf{q}}\pm\tilde{g}^\nu_{\mathbf{k}}$, and $\nu, \nu'=x,y$. 

While the model in Eq.~(\ref{eq:RKKYHam}) includes complicated interactions, the dominant contributions can be extracted when $\chi^0_{\mathbf{k}\sigma\sigma'}$ 
shows conspicuous peaks in the Brillouin zone. 
This often happens when the Fermi surface is partially nested. 
In this case, the sum of the wave numbers in Eq.~(\ref{eq:RKKYHam}) is dominated by the contributions from the peak wave numbers. 
In the case of the square lattice, there appear two or four pairs of peaks, $\mathbf{Q}_\eta$ ($\eta=1,2$ or $1$-$4$), which are orthogonal to each other, reflecting the fourfold rotational symmetry of the point group $C_{4v}$. 
In the absence of the SOC ($\mathcal{H}^{\rm ani}=\mathcal{H}^{\rm DM}=0$), the system with such conspicuous peaks in $\chi^0_{\mathbf{k}\sigma\sigma'}$ exhibits a noncoplanar spin texture characterized by the two wave numbers, while it is not a skyrmion type~\cite{Ozawa_doi:10.7566/JPSJ.85.103703,Hayami_PhysRevB.95.224424}. 
The SOC is expected to increase a chance for further exotic spin textures including skyrmions through the additional contributions in $\mathcal{H}^{\rm ani}$ and $\mathcal{H}^{\rm DM}$.  

Then, we end up with the effective spin model, whose Hamiltonian is summarized as 
\begin{eqnarray}
\label{eq:Hameff}
\mathcal{H} = &-&2\sum_{\eta}\Big( \sum_{\alpha \beta}
J_{\eta}^{\alpha \beta} S^{\alpha}_{\mathbf{Q}_{\eta}} S^{\beta}_{-\mathbf{Q}_{\eta}}+i \mathbf{D}_{\eta} \cdot (\mathbf{S}_{\mathbf{Q}_{\eta}} \times \mathbf{S}_{-\mathbf{Q}_\eta})\Big)  \nonumber \\
&-& H \sum_i S_i^z, 
\end{eqnarray}
where the sum of $\eta$ is taken for the set of $\mathbf{Q}_\eta$ giving the multiple maxima in the bare susceptibility $\chi^0_{\mathbf{k}\sigma\sigma'}$. 
$J^{\alpha\beta}_{\eta}$ and $\mathbf{D}_{\eta}$ are the coupling constants for symmetric and antisymmetric exchange interactions in momentum space ($\alpha, \beta=x, y, z$), which are long-ranged in real space. 
These exchange couplings are related to the coefficients in Eqs.~(\ref{eq:RKKYHam_iso})-(\ref{eq:RKKYHam_DM}) and $\chi^0_{\mathbf{k}\sigma\sigma'}$. 
In Eq.~(\ref{eq:Hameff}), we added the Zeeman coupling to an external magnetic field in the $z$ direction. 

The number of independent coupling constants in the model in Eq.~(\ref{eq:Hameff}) depends on the set of $\mathbf{Q}_\eta$. 
In the present study, we consider the case with two $\mathbf{Q}_\eta$ as $\mathbf{Q}_1=(0,Q^*)$ and $\mathbf{Q}_2=(Q^*,0)$, which leads to four independent coupling constants, $J^{xx}$, $J^{yy}$, $J^{zz}$, and $D$: $J_1^{xx}=J_2^{yy}\equiv J^{xx}$, $J_1^{yy}=J_2^{xx}\equiv J^{yy}$, $J_1^{zz}=J_2^{zz} \equiv J^{zz}$, and $D_{1}^x=D_2^y \equiv D$ (all other components are zero). 
The extension to other cases is straightforward. 
In the following calculations, we take $Q^*=\pi/4$ and set the energy unit as $J^{xx}+J^{yy}+J^{zz}=1$ without loss of generality. 

We investigate the magnetic phase diagram of the effective spin model in Eq.~(\ref{eq:Hameff}) on the square lattice by performing simulated annealing from high temperature~\cite{comment_simulation}. 
Our simulations are carried out with the standard Metropolis local updates under periodic boundary conditions in both directions. 
We present the results for the system with $N=48^2$ sites. 
In the simulation, we first perform simulated annealing to find the low-energy configuration by gradually reducing the temperature with the rate $T_{n+1}=\alpha T_{n}$, where $T_{n}$ is the temperature in the $n$th step. 
We set the initial temperature $T_0 = 10^{-1}$-$10^{0}$ and take the coefficient $\alpha=0.9995$-$0.9999$. 
The final temperature, which is typically taken at $T=10^{-4}$, is reached after $10^5$-$10^6$ Monte Carlo sweeps in total.

\begin{figure}[t]
\begin{center}
\includegraphics[width=1.0 \hsize]{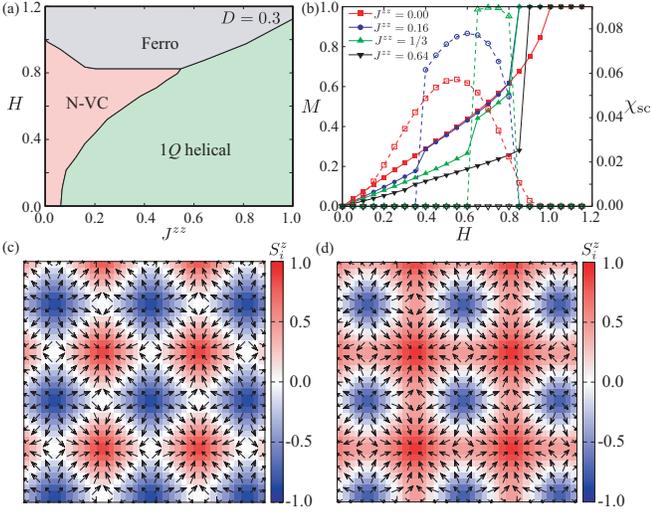} 
\caption{
\label{Fig:zzline}
(a) Magnetic phase diagram while changing $J^{zz}$ and $H$ at $D=0.3$ and $J^{xx}=J^{yy}$ ($J^{xx}+J^{yy}+J^{zz}=1$). 
$1Q$ helical, N-VC, and Ferro represent the single-$Q$ helical, the N\'eel-type vortex crystal, and the forced ferromagnetic states, respectively. 
(b) Magnetization curves denoted by filled symbols (left axis) and the net scalar chirality denoted by open symbols (right axis) for $J^{zz}=0$, 0.16, 1/3, and 0.64. 
(c), (d) Snapshots of the spin configurations in the N-VC phase for $J^{zz}=0$ at (c) $H=0$ and (d) $H=0.3$. 
The arrows and contour denote the $xy$ and $z$ components of the spin moments, respectively. 
}
\end{center}
\end{figure}

Figure~\ref{Fig:zzline}(a) shows the $J^{zz}$-$H$ phase diagram for the model in Eq.~(\ref{eq:Hameff}) obtained by the simulated annealing for $J^{xx}=J^{yy}$ and $D=0.3$. 
We find three phases in the parameter range, which are distinguished by computing the spin structure factor defined as $S_s^{\alpha \alpha}(\mathbf{q}) 
= (1/N) \sum_{j,l} \langle S_j^{\alpha} S_l^{\alpha} \rangle e^{i \mathbf{q}\cdot (\mathbf{r}_j-\mathbf{r}_l)}$,
where $\alpha=x,y,z$. 
We also compute the magnetization $M=\langle(1/N)\sum_i S^{z}_i\rangle$ and 
the uniform scalar chirality defined as $\chi_{\rm sc} = \langle (1/N) \sum_{j, \delta = \pm1} \mathbf{S}_j \cdot (\mathbf{S}_{j+\delta\hat{x}} \times \mathbf{S}_{j+\delta \hat{y}})\rangle$, 
where $\hat{x}$ and $\hat{y}$ are the unit translation vectors in the $x$ and $y$ direction, respectively~\cite{Yi_PhysRevB.80.054416}. 

The single-$Q$ (1$Q$) helical state appears for small $H$ and large $J^{zz}$. 
It has a cycloidal structure, e.g., 
$\mathbf{S}_i\propto [0, a^y \cos (\mathbf{Q}_1 \cdot \mathbf{r}_i), a^z \sin (\mathbf{Q}_1 \cdot \mathbf{r}_i)]$ or 
$\mathbf{S}_i\propto [a^x \cos (\mathbf{Q}_2 \cdot \mathbf{r}_i), 0, a^z \sin (\mathbf{Q}_2 \cdot \mathbf{r}_i)]$. 
Note that this is not a simple single-$Q$ order: the spiral is not circular but elliptic with the coefficients $a^z>a^x$, $a^y$ ($a^z<a^x$, $a^y$) for $J^{zz}>J^{xx}$, $J^{yy}$ ($J^{zz}<J^{xx}$, $J^{yy}$). 

Meanwhile, in the smaller $J^{zz}$ region, the system shows a periodic array of magnetic vortices. 
This spin texture is characterized by two wave numbers (double-$Q$): the spin structure factor has the peaks at $\mathbf{q}=\pm\mathbf{Q}_1$ and $\pm\mathbf{Q}_2$, in addition to the uniform component at $\mathbf{q}=\mathbf{0}$ for $H \neq 0$. 
The real-space spin configuration is 
described as the equal superposition of two helices,  
\begin{equation}
\label{Eq:NVC}
\mathbf{S}_i \propto (\cos \mathcal{Q}_2, \cos \mathcal{Q}_1, a^z ( \sin \mathcal{Q}_1 +\sin \mathcal{Q}_2)+ b),
\end{equation}
where $\mathcal{Q}_\eta=\mathbf{Q}_\eta \cdot \mathbf{r}_i+|\mathbf{Q}_\eta|/2 $; the coefficients $a^z$ and $b$ depend on $J^{zz}$ and $H$. 
Figures~\ref{Fig:zzline}(c) and \ref{Fig:zzline}(d) show typical real-space spin configurations obtained from a snapshot of the simulated annealing at $H=0$ and 0.3 for $J^{zz}=0$, respectively. 
At $H=0$, the spin configuration consists of a periodic array of vortices with the positive $S^z_i$ (helicity $\pi$) and antivortices with the negative $S^z_i$ (helicity $0$), both of which are of N\'eel type [see Fig.~\ref{Fig:ponti}(b)]. 
Thus, this is regarded as a N\'eel-type vortex crystal (N-VC). 
In the N-VC state, although both vortex and antivortex carry nonzero scalar chirality, the net value vanishes due to the cancellation between them. 
With an increase of $H$, the vortices extend and the antivortices shrink, which leads to a skyrmion-like crystal of N\'eel type, as shown in Fig.~\ref{Fig:zzline}(d)~\cite{comment_skyrmion}. 
In this region, $\chi_{\rm sc}$ becomes nonzero, as plotted in Fig.~\ref{Fig:zzline}(b). 
While further increasing $H$, the spins are overall aligned along the field direction, and hence, $\chi_{\rm sc}$ is suppressed as shown in Fig.~\ref{Fig:zzline}(b). 

The remarkable point is that the N-VC phase is stable even in the absence of the magnetic field. 
As mentioned in the introduction, although a N\'eel-type skyrmion is expected for the Rashba SOC, it is usually stabilized in an applied magnetic field. 
Our results indicate that itinerant nature of electrons increases a chance for the N\'eel-type spin textures. 

We plot the magnetization curves at several $J^{zz}$ in Fig.~\ref{Fig:zzline}(b). 
In the region where the N-VC phase is stabilized at $H=0$ for small $J^{zz}$ ($J^{zz}\lesssim 0.06$), the magnetization continuously increases with $H$ and smoothly approaches the saturation in the forced ferromagnetic state at $H \sim 1$. 
Meanwhile, the magnetization jumps at the phase boundaries between the N-VC and the ferromagnetic state as well as between N-VC and 1$Q$ helical, as shown in Fig.~\ref{Fig:zzline}(b) for $J^{zz}=0.16$ and $1/3$. 
We note that this behavior is similar to the case in the Heisenberg model with the short-ranged DM interaction~\cite{Mochizuki_PhysRevLett.108.017601,Banerjee_PhysRevX.4.031045,Rowland_PhysRevB.93.020404}. 
For larger $J^{zz}=0.64$, there is a magnetization jump between the 1$Q$ helical and ferromagnetic states.

\begin{figure}[t]
\begin{center}
\includegraphics[width=1.0 \hsize]{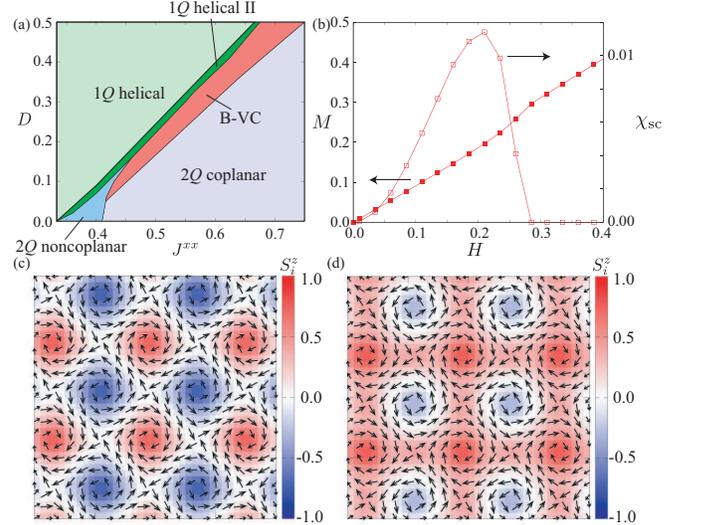} 
\caption{
\label{Fig:xxline}
(a) Phase diagram in the $J^{xx}$-$D$ plane at $H=0$ for $J^{yy}=J^{zz}$. 
B-VC represents the Bloch-type vortex crystal state. 
(b) Magnetization curve denoted as filled symbols (left axis) and the net scalar chirality denoted as open symbols (right axis) for $J^{xx}=0.5625$ and $D=0.3$. 
(c), (d) Snapshots of the spin configurations in the B-VC phase for $J^{xx}=0.5625$ and $D=0.3$ at (c) $H=0$ and (d) $H=0.235$.  
The arrows and contour denote the $xy$ and $z$ components of the spin moments, respectively. 
}
\end{center}
\end{figure}

Next, let us discuss another type of VCs found in the model in Eq.~(\ref{eq:Hameff}) with different anisotropy. 
Figure~\ref{Fig:xxline}(a) shows the $J^{xx}$-$D$ phase diagram at $H=0$ and $J^{yy}=J^{zz}$ obtained from the simulated annealing. 
There are additional four phases besides the 1$Q$ helical state: 1$Q$ helical II, Bloch-type vortex crystal (B-VC), double-$Q$ (2$Q$) coplanar, and 2$Q$ noncoplanar states. 

The upper-left region of the phase diagram is dominated by the 1$Q$ helical states. 
The difference between the 1$Q$ helical and 1$Q$ helical II states lies in the $x$ ($y$) component of the spin configuration for the ordering vector $\mathbf{Q}_1$ ($\mathbf{Q}_2$): it becomes nonzero in the 1$Q$ helical II state. 
On the other hand, the lower-right region is occupied by the double-$Q$ (2$Q$) states. 
The spin configuration in the 2$Q$ coplanar state is expressed as $\mathbf{S}_i \propto [\cos (\mathcal{Q}_1+\pi), \cos \mathcal{Q}_2, 0]$. 
In the small $J^{xx}$ and $D$ region, the system shows a 2$Q$ noncoplanar order, whose spin configuration is given by $\mathbf{S}_i \propto [a^x\cos (\mathcal{Q}_1+ \pi)+ a'^x \cos \mathcal{Q}_2, a'^y \cos \mathcal{Q}_2+ a^y \cos (\mathcal{Q}_1+\pi), a^z \sin \mathcal{Q}_2]$ where $a^\mu$ and $a'^{\mu}$ take different values. 
The net scalar chirality is zero in both 2$Q$ states, although the latter state accompanies a chiral density wave modulated with $\mathbf{Q}_1$. 

In the competing region between the 1$Q$ and 2$Q$ states, we find that the system exhibits another VC. 
The real-space spin configuration is approximately given by 
\begin{eqnarray}
\label{Eq:BVC}
\mathbf{S}_i \propto
\left(
    \begin{array}{c}
    \cos (\mathcal{Q}_1+ \pi)+ 
    a^x \cos \mathcal{Q}_2 \\
    \cos \mathcal{Q}_2+ a^x \cos (\mathcal{Q}_1+\pi) \\    
    a^z ( \sin \mathcal{Q}_1 +\sin \mathcal{Q}_2)
   \end{array}
  \right)^{\rm T},  
\end{eqnarray}
where $a^x<1$ and ${\rm T}$ denotes the transpose. 
While this double-$Q$ state has a similar periodic array of noncoplanar spin textures with positive and negative $S_i^z$ (helicity $\simeq-\pi/2$ and $\simeq\pi/2$, respectively), it has a distinct aspect from the N-VC state in Eq.~(\ref{Eq:NVC}): the spins rotate in the tangential planes when moving from core to periphery, as shown in Fig.~\ref{Fig:xxline}(c). 
This swirling texture is similar to the Bloch-type skyrmion in Fig.~\ref{Fig:ponti}(a), and hence, we call this state the Bloch-type vortex crystal (B-VC). 

This is, to the best of our knowledge, the first example of the stable B-VC state without the Dresselhaus SOC. 
It is also surprising that the B-VC state is stable even at zero field, similar to the N-VC state. 
From our results, the emergence of B-VC is understood from the competition between the anisotropic symmetric exchange and the antisymmetric exchange.  
The former tends to favor the 2$Q$ coplanar state discussed above, while the latter tends to favor the 1$Q$ cycloidal state or the N-VC in Eq.~(\ref{Eq:NVC}). 
Indeed, the spin configuration of B-VC in Eq.~(\ref{Eq:BVC}) is regarded as a superposition of the 2$Q$ and N-VC textures. 

Finally, let us study the effect of the magnetic field on the B-VC. 
We plot the magnetization and the scalar chirality at $J^{xx}=0.5625$ and $D=0.3$ in Fig.~\ref{Fig:xxline}(b). 
Both quantities become nonzero for $H>0$. 
This is because the B-VC state evolves into a skyrmion-like crystal while extending the positive-$S_i^z$ region, as shown in Fig.~\ref{Fig:xxline}(d). 
With a further increase of $H$, the B-VC phase is replaced by the 2$Q$ coplanar state with an additional ferromagnetic component induced by the magnetic field for $H\gtrsim 0.285$. 
At the same time, the scalar chirality vanishes. 
There is a small jump in the magnetization curve suggesting a discontinuous phase transition, although a more careful analysis is required to settle this point. 

To summarize, we have theoretically shown that the interplay between the SOC and the spin-charge coupling in itinerant magnets gives rise to fertile possibilities of exotic magnetic orderings. 
Starting from the Kondo lattice model with the Rashba SOC for polar magnetic conductors, we derived an effective spin model with long-ranged anisotropic and antisymmetric exchange interactions. 
By simulated annealing, we found that the model exhibits VCs of both N\'eel and Bloch type even in the absence of a magnetic field. 
The latter Bloch-type one is rather surprising, as it is usually induced by the Dresselhaus SOC. 
In an applied magnetic field, we showed that both VCs turn into skyrmion-like textures.

Our results indicate that a variety of noncoplanar spin textures of vortex and skrymion types can be induced and controlled by the form of the SOC and the electronic dispersion. 
The SOC would be designed by the surface and heterostructures, and also controlled by an external electric field. 
The electronic dispersion is inherit to the materials, while it would be modulated by an external pressure and a strain in the heterostructures. 
Thus, we believe that our results are useful not only to give an insight into the origin of complex noncoplanar spin structures observed in experiments, e.g., for monolayer metals on substrates~\cite{Bergmann_PhysRevLett.96.167203,heinze2011spontaneous,Yoshida_PhysRevLett.108.087205,Zimmermann_PhysRevB.90.115427,Serrate_PhysRevB.93.125424,hoffmann2017antiskyrmions}, but also to pave the way for exploring further exotic spin textures in polar and chiral systems. 

\begin{acknowledgments}
This research was supported by JSPS KAKENHI Grants Numbers 15K05176 and 16H06590. Parts of the numerical calculations were performed in the supercomputing systems in ISSP, the University of Tokyo.
\end{acknowledgments}
\bibliographystyle{apsrev}
\bibliography{ref}

\begin{thebibliography}{59}
\bibitem[{com({\natexlab{a}})}]{comment_simulation}
\bibinfo{note}{We also confirm that magnetic phases are obtained by classical
  Monte Carlo simulations at temperatures $\lesssim 10^{-2}$.}
\bibitem[{com({\natexlab{b}})}]{comment_skyrmion}
\bibinfo{note}{The skyrmion core is located at the center of the square
  plaquette in the lattice structure.}
\end{thebibliography}

\end{document}